\documentclass[10pt,conference]{IEEEtran}\IEEEoverridecommandlockouts
\usepackage{epsfig,graphicx,subfigure,psfrag,amsmath,amsthm,cases}
\usepackage{latexsym,amssymb,epsfig,subfigure,algorithm,mathtools}
\usepackage{algorithmic}
\usepackage{algorithm}
\usepackage{color}
\usepackage{url}
\usepackage{cite}
\usepackage{scrtime}
\usepackage{bm}
\usepackage[english]{babel}

\date{\thistime,\,\today}

\newtheorem{theo}{Theorem}

\newtheorem{rema}{Remark}

\newcommand*{\QEDA}{\hfill\ensuremath{\blacksquare}}
\DeclareMathOperator{\sinc}{sinc}
\begin{document}

\title{Mitigating Pilot Contamination in Multi-cell Hybrid Millimeter Wave Systems}
\author{\thanks{The authors are with $^{\dag}$the School of Electrical Engineering and Telecommunications, UNSW, and $^{\ddag}$the School of Engineering and Information Technology, UNSW (Email: \{lou.zhao, zhiqiang.wei\}@student.unsw.edu.au, \{w.k.ng, j.yuan, mark.reed\}@unsw.edu.au). D. W. K. Ng is supported under the Australian Research Council (ARC) Discovery Early Career Researcher Award funding scheme DE170100137. Jinhong Yuan is currently on leave from UNSW and with CAS. This work was supported in part by the ARC Discovery Project DP160104566, Linkage Project LP160100708, and CAS Pioneer Hundred Talents Program.}
\IEEEauthorblockN{Lou~Zhao$^{\dag}$,  Zhiqiang~Wei$^{\dag}$, Derrick~Wing~Kwan~Ng$^{\dag}$,  Jinhong Yuan$^{\dag}$, and Mark~C.~Reed$^{\ddag}$}
}

\maketitle

\begin{abstract}
In this paper, we investigate the system performance of a multi-cell multi-user (MU) hybrid millimeter wave (mmWave) multiple-input multiple-output (MIMO) network adopting the channel estimation algorithm proposed in \cite{Zhao2017} for channel estimation.
Due to the reuse of orthogonal pilot symbols among different cells, the channel estimation is expected to be affected by pilot contamination, which is considered as a fundamental performance bottleneck of conventional multi-cell MU massive MIMO networks.
To analyze the impact of pilot contamination on the system performance, we derive the closed-form approximation expression of the normalized mean squared error (MSE) of the channel estimation performance.
Our analytical and simulation results show that the channel estimation error incurred by the impact of pilot contamination and noise vanishes asymptotically with an increasing number of antennas equipped at each radio frequency (RF) chain deployed at the desired BS.
Thus, pilot contamination is no longer the fundamental problem for multi-cell hybrid mmWave systems.
\end{abstract}\vspace*{-0mm}
\vspace*{-0mm}
\section{Introduction}

Recently, the escalating demands for high data transmission, which is one of the key requirements of the fifth-generation (5G) wireless communication systems, have triggered and attracted tremendous interests from both academia and industry, e.g. \cite{Kokshoorn2016,Dai2013,AZhang2015,Lin2017,Akbar2017,Ng2017,Marzetta2010,Wei2017,Zhao2008}.
To meet the ultra-high data rate requirement of emerging applications, millimeter wave (mmWave) massive multiple-input multiple-output (MIMO) systems have been proposed \cite{Zhao2017B,Bjornson2016}.
In practice, to strike a balance between data rate, hardware cost\footnote{The hardware cost is mainly contributed by analog to digital converter/digital to analog converter (ADC/DAC) associated with each RF chain.}, system complexity, and power consumption, hybrid mmWave MIMO systems are proposed for practical implementation \cite{Ng2017,Bjornson2016}.
In particular, for hybrid MIMO architecture, the required numbers of radio frequency (RF) chains equipped at a base station (BS) and users are much smaller than the numbers of antennas equipped at the BS.
Thus, both the hardware cost and the energy consumption of hybrid mmWave massive MIMO systems, can be reduced significantly compared to the conventional fully digital mmWave massive MIMO architecture, e.g. \cite{Ng2017,AZhang2015}.
Furthermore, the severe propagation path loss of mmWave channels between the transceivers can be compensated by forming a highly directional information beam enabled by the massive number of antennas.

In practice, the performance of hybrid mmWave systems depends on the accuracy of estimated channel state informnation (CSI).
However, channel estimation algorithms for hybrid mmWave systems are different from that of fully digital systems due to the hardware limitation in the former systems.
Currently, majority of contributions in the literature focus on the development of CSI feedback based channel estimation methods for frequency-division duplex (FDD) hybrid mmWave systems, e.g. \cite{Alkhateeb2015}.
Mainly, it is motivated by the assumption of sparsity of mmWave channels that the numbers of resolvable angles of arrival (AoA)/departure (AoD) paths are finite and limited.
Thus, CSI acquisition via feedbacks only leads to a small amount of signaling overhead compared to non-sparse CSI acquisition.
However, in some scenarios, the assumption of the sparsity of mmWave channel may not hold.
For example, for practical urban micro-cell (UMi) scenarios, such as the city center, the number of scattering clusters increases significantly and the channels are expected to be non-sparse.
In \cite{Akdeniz2014}, recent field test results as well as ray-tracing simulation results have shown that reflections from street signs, lamp posts, parked vehicles, passing people, etc., could reach a receiver from all possible directions in UMi scenarios.
Besides, recent field measurements also confirm that both LOS components and scattering components exist in the inter-cell mmWave channels in small-cell systems \cite{Akdeniz2014}.
Thus, Rician fading channel model is more suitable for modeling the inter-cell channels in small-cell systems.
Under such circumstances, the signaling overhead may be too large to be acquired via feedback.
To overcome the aforementioned common drawbacks of conventional CSI feedback based FDD mmWave channel estimation algorithms, a novel TDD-based beamforming channel estimation algorithm, which exploits orthogonal pilot symbols transmission from users to the BS via the strongest received line-of-sight (LOS) paths, was proposed in \cite{Zhao2017}.
It was shown that the proposed algorithm for hybrid systems can achieve a considerable achievable rate of the optimal fully digital systems and possess robustness against the hardware imperfection.

Despite the channel estimation algorithm proposed in \cite{Zhao2017} offers a viable solution to unlock the potential of mmWave systems, it only considered a simple single-cell scenario.
In practice, multi-cell systems are usually deployed to improve the spectral efficiency.
Besides, it is expected that small-cell will serve as a core structure of future cellular systems \cite{Ngo2017}.
However, due to the short radius in small-cell, mmWave communication systems may suffer severe inter-cell interferences during uplink channel estimation and downlink transmission.
Since the resources of orthogonal pilot are limited, they are reused among different cells for multi-cell channel estimation.
In this case, the received pilot symbols from the users in the desired cell for channel estimation are affected by the reused pilot symbols from the users in neighboring cells, which is known as pilot contamination \cite{Marzetta2010}.
In fact, pilot contamination is considered as a fundamental performance bottleneck of the conventional multi-cell multiuser (MU) massive MIMO systems, since the resulting channel estimation errors do not vanish even if the number of antennas is sufficiently large, cf. \cite{Marzetta2010}.
In the literature, most of the existing works for multi-cell massive MU-MIMO systems modeled the inter-cell uplink channels as Rayleigh fading channels, e.g. \cite{Marzetta2010,Jose2011}.
However, existing analysis cannot be applied to small-cell systems directly due to the presence of strong LOS inter-cell channels.
Furthermore, a thorough study on the impact of pilot contamination in such practical systems has not been reported yet.

Motivated by the aforementioned discussions, we consider a multi-cell MU hybrid mmWave system.
In particular, we apply the non-feedback TDD-based mmWave channel estimation algorithm proposed in \cite{Zhao2017} to a multi-cell scenario and study the corresponding performance of channel estimation under the impact of pilot contamination.
Analysis and simulation results reveal that the channel estimation error caused by the impact of pilot contamination would vanish asymptotically with the increasing number of antennas.
Our main contributions are summarized as follows:

\begin{itemize}

\item Applying the three-step mmWave channel estimation algorithm proposed in \cite{Zhao2017} to a multi-cell scenario, we study the impact of pilot contamination on the uplink mmWave channel estimation due to the reuse of orthogonal pilot symbols among different cells.
Our results reveal that in the phase of channel estimation, the analog receive beamforming matrix adopted at the desired BS forms a spatial filter which blocks the signal reception of the undesired pilot symbols from neighboring cells. In particular, with an increasing number of antennas equipped at each RF chain, the mainlobe beamwidth of the spatial filter, which aligns to strongest AoA directions, becomes narrower and the amplitude of sidelobes becomes lower.
Thus, the impact of pilot contamination caused by the users outside strongest AoA directions can be mitigated. We mathematically prove that the normalized mean square error (MSE) performance of the channel estimation algorithm proposed in \cite{Zhao2017} improves proportionally with the increasing number of antennas equipped at each RF chain, which is different from previous results in \cite{Marzetta2010,Jose2011}.

\item We adopt zero-forcing (ZF) precoding for the downlink transmission based on the estimated CSI. The simulation results show that the average achievable rate per user under the impact of pilot contamination increases with the increasing number of antennas equipped at the BS.

\end{itemize}
Notation: $\mathbb{E}_{\mathrm{h}}(\cdot )$ denotes statistical expectation operation with respect to random variable $h$, $\mathbb{C}^{M\times N}$ denotes the space of all $M\times N$ matrices with complex entries; $(\cdot )^{-1}$ denotes inverse operation; $(\cdot )^{H}$ denotes Hermitian transpose; $(\cdot )^{\ast }$ denotes complex conjugate; $(\cdot)^{T}$ denotes transpose; $|\cdot |$ denote the absolute value of a complex scalar; $\mathrm{tr}(\cdot )$ denotes trace operation; $\mathrm{sinc}(x)$ denotes a sinc function with input $x$, i.e., $\frac{\sin\left(x\right)}{x}$.
The distribution of a circularly symmetric complex Gaussian (CSCG) random vector with a mean vector $\mathbf{x}$ and a covariance matrix ${\sigma}^{2}\mathbf{I}$  is denoted by ${\cal CN}(\mathbf{x},{\sigma}^{2}\mathbf{I})$, and $\sim$ means ``distributed as". $\mathbf{I}_{\mathrm{P}}$ is an $P \times P$ identity matrix.

\section{System Model}
In this paper, a multi-cell MU hybrid subarray mmWave system is considered. The system consists of $L$ neighboring cells and there are one BS and $N$ users in each cell, cf. Fig. $1$.
The BS in each cell is equipped with $N_{\mathrm{RF}}$ RF chains serving the $N$ users simultaneously.
We assume that each RF chain equipped at the BS can access to an uniform linear array (ULA) with $M$ antennas by using $M$ phase shifters.
Besides, each user is equipped with one RF chain and a $P$-antenna array.
In addition, we focus on $M\geqslant N_{\mathrm{RF}}$, which exploits large antenna array gain with limited number of RF chains.
To simplify the analysis in the following sections, without loss of generality, we set $N_{\mathrm{RF}} = N$ and each cell has the same number of RF chains at the BS.

According to the widely adopted setting for multi-cell TDD in uplink channel estimation and downlink data transmission, we assume that the users and the BSs in all cells are fully synchronized in time \cite{Marzetta2010,Jose2011}.
We denote $\mathbf{H}_{k}\in\mathbb{C}^{M\times P}$ as the uplink channel matrix between the desired BS and user $k$ in the desired cell.
Besides, $\mathbf{H}_{k}$ is a narrowband slow time-varying block fading channel.
\begin{figure}[t]\vspace{-0mm}
\centering
\includegraphics[width=2.5in]{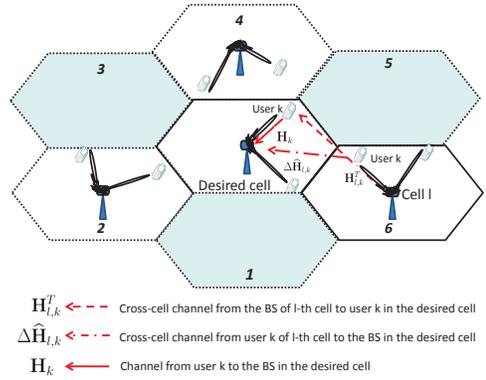}
\vspace{-3mm}
\centering
\caption{A multi-cell MU mmWave cellular system with $L=6$ neighboring cells.}
\label{fig:MCM}\vspace{-4mm}
\end{figure}
Recent field tests show that both strong LOS components and non-negligible scattering components may exist in mmWave propagation channels \cite{Akdeniz2014}, especially in the urban areas.
Therefore, mmWave channels can also be modeled by non-sparse Rician fading with a large Rician K-factor \cite{Akdeniz2014}.
In this paper, we assume that the uplink channel $\mathbf{H}_k$ consists of a deterministic LOS channel matrix $\mathbf{H}_{\mathrm{L},k}\in\mathbb{C}^{M\times P}$ and a scattered channel matrix $\mathbf{H}_{\mathrm{S,}k}\in\mathbb{C}^{M\times P}$, which is given by: \vspace*{-1mm}
\begin{equation}\label{eqn:LOS_channel}
\mathbf{H}_{k}=\underset{\mathrm{LOS}}{\underbrace{\mathbf{H}_{\mathrm{L,}k}\mathbf{G}_{\mathrm{L,}k}}}+\underset{\mathrm{Scattering}}{\underbrace{\mathbf{H}_{\mathrm{S,}k}\mathbf{G}_{\mathrm{S,}k}}},\vspace*{-1mm}
\end{equation}%
where $\mathbf{G}_{\mathrm{L,}k}\in\mathbb{C}^{P\times P}$ and $\mathbf{G}_{\mathrm{S},k}\in\mathbb{C}^{P\times P}$ are diagonal matrices as\vspace*{-1mm}
\begin{equation}
\mathbf{G}_{\mathrm{L,}k}= \sqrt{\frac{\upsilon _{k}}{\upsilon _{k}+1}}\mathbf{I}_{\mathrm{P}} \quad \mbox{and}\quad \ \mathbf{G}_{\mathrm{S,}k}= \sqrt{\frac{1}{\upsilon _{k}+1}}\mathbf{I}_{\mathrm{P}},\label{Eq:2}\vspace*{-1mm}
\end{equation}%
respectively, and $\upsilon _{k}>0$ is the Rician K-factor of user $k$.
We assume that AoAs from all the users to the desired BS are various.
In addition, all the users are separated by at least hundreds of wavelengths\footnote{For a carrier frequency of $30$ GHz, the distance of one hundred wavelengths is approximately $1$ meter.} and we can assume that their channels to the desired/neighboring BS are uncorrelated \cite{Marzetta2010}.
Thus, we can express the deterministic LOS channel matrix $\mathbf{H}_{\mathrm{L},k}$ of user $k$ in the desired cell as \cite{book:wireless_comm}\vspace*{-0.0mm}
\begin{equation}
\mathbf{H}_{\mathrm{L,}k}=\sqrt{\varpi_{{k}}}\mathbf{h}_{\mathrm{L,}k}^{\mathrm{BS}}\mathbf{h}_{%
\mathrm{L,}k}^{H},\label{Eq:3}
\end{equation}\vspace*{-0.0mm}%
where $\varpi_{{k}}$ accounts for the corresponding large scale path loss, $\mathbf{h}_{\mathrm{L},k}^{\mathrm{BS}}$ $\in\mathbb{C}^{M\times 1}$ and $\mathbf{h}_{\mathrm{L,}k}$ $\in\mathbb{C}^{P\times 1}$ are the antenna array response vectors of the BS and user $k$, respectively.



In particular, $\mathbf{h}_{\mathrm{L,}k}^{\mathrm{BS}}$ and $\mathbf{h}_{\mathrm{L,}k}$ can be expressed as \cite{book:wireless_comm}\vspace*{-0mm}
\begin{align}
\mathbf{h}_{\mathrm{L},k}^{\mathrm{BS}}& =\left[
\begin{array}{ccc}
1,  \ldots
,e^{-j2\pi \left( M-1\right) \tfrac{d}{\lambda }\cos \left(
\theta _{k}\right) }%
\end{array}\label{Eq:4}
\right] ^{T}\text{ and} \\
\mathbf{h}_{\mathrm{L},k}& =\left[
\begin{array}{ccc}
1,  \ldots ,
e^{-j2\pi \left( P-1\right) \tfrac{d}{\lambda }\cos \left( \phi
_{k}\right) }%
\end{array}%
\right] ^{T}, \label{Eq:5}
\end{align}\vspace*{-0mm}%
respectively, where $d$ is the distance between the neighboring antennas at the BS and users and $\lambda $ is the wavelength of the carrier frequency.
Variables $\theta _{k}\in \left[ 0,\text{\ }\pi \right]$ and $\phi _{k}\in \left[ 0,\text{\ }\pi \right] $ are the angles of incidence of the LOS path at antenna arrays of the desired BS and user $k$, respectively.
As commonly adopted in the literature \cite{book:wireless_comm}, we set $d=\frac{\lambda }{2}$ for convenience.
Without loss of generality, we assume that the scattering component $\mathbf{H}_{\mathrm{S,}k}$ consists of $N_{\mathrm{cl}}$ propagation paths, which can be expressed as\vspace*{-0mm}
\begin{align}
\vspace*{-0mm}
\mathbf{H}_{k}&=\sqrt{\varpi_{{k}}}\sqrt{\tfrac{1}{{N_{\mathrm{cl}}}}}\overset{N_{\mathrm{cl}}}{\underset{i=1}{\sum }}{\alpha _{i}}\mathbf{h}_{i}^{\mathrm{BS}}\mathbf{h}_{k,i}^{H},\label{Eq:6}
\end{align}\vspace*{-0mm}
where $\mathbf{h}_{i}^{\mathrm{BS}}\in\mathbb{C}^{M\times 1}$ and $\mathbf{h}_{k,i}\in\mathbb{C}^{P\times 1}$ are the antenna array response vectors of the BS and user $k$ associated to the $i $-th NLOS propagation path, respectively.
Here $\alpha _{i}\sim \mathcal{CN}\left( 0,1\right) $ represents the path attenuation of the $i$-th NLOS propagation path and $\mathbf{h}_{\mathrm{S},k}\in\mathbb{C}^{M\times 1}$ is the $k$-th column vector of $\mathbf{H}_{\mathrm{S},k}$.
With the increasing number of clusters, the path attenuation coefficients and the AoAs between the users and the BS behave randomly.
Let $\Delta\widehat{\mathbf{H}}_{l,k}\in\mathbb{C}^{M\times P}$ be the inter-cell mmWave uplink channel between user $k$ in the $l$-th neighboring cell and the desired BS, $\forall l\in \left\{ 1,\ldots, L\right\}$, cf. Fig. \ref{fig:MCM}.
Since the inter-site distance is short in small-cell systems, the inter-cell channels usually contain strong LOS components\footnote{The probability of the existence of the LOS components decreases with increasing the radius of a cell.} \cite{Akdeniz2014}.
Thus, the inter-cell uplink mmWave channels can be expressed as
\begin{align}
\hspace{-3mm}\Delta\widehat{\mathbf{H}}_{l,k}&=\sqrt{\widehat{\varpi}_{l,{k}}}\underset{\mathrm{LOS}}{\underbrace{\Delta\widehat{\mathbf{H}}_{\mathrm{L,}l,k}\widehat{\mathbf{G}}_{\mathrm{L,}l,k}}}+\sqrt{\widehat{\varpi}_{l,{k}}}\underset{\mathrm{Scattering}}{\underbrace{\Delta\widehat{\mathbf{H}}_{\mathrm{S,}l,k}\widehat{\mathbf{G}}_{\mathrm{S,}l,k}}},
\end{align}
where $\widehat{\varpi}_{l,{k}}$ is the corresponding large scale path loss coefficient.
We note that inter-cell Rician K-factor matices $\widehat{\mathbf{G}}_{\mathrm{L,}l,k}= \sqrt{\frac{\upsilon _{l,k}}{\upsilon _{l,k}+1}}\mathbf{I}_{\mathrm{P}}$, $\widehat{\mathbf{G}}_{\mathrm{S,}l,k}= \sqrt{\frac{1}{\upsilon _{l,k}+1}}\mathbf{I}_{\mathrm{P}}$, LOS components
$\Delta\widehat{\mathbf{H}}_{\mathrm{L,}l,k}=\Delta\widehat{\mathbf{h}}_{\mathrm{L,}l,k}^{\mathrm{BS}}\Delta\widehat{\mathbf{h}}_{%
\mathrm{L,}l,k}^{H}$ follows similar assumptions as in expression (\ref{Eq:3})$-$(\ref{Eq:5}).
Besides, we adopt a similar assumption as in expression (\ref{Eq:6}) for the scattering components $\Delta\widehat{\mathbf{H}}_{\mathrm{S,}l,k}$.
All these mentioned inter-cell propagation path loss coefficients are modeled as in \cite{Akdeniz2014}.
According to recent field measurements, e.g. \cite{Akdeniz2014}, the typical values of Rician K-factors $\upsilon_{l,k}$ for $\Delta\widehat{\mathbf{H}}_{l,k}$ is $\upsilon_{l,k}\in[0, \text{\ }5]$.

\section{Multi-cell Uplink Channel Estimation Performance Analysis}

\begin{figure}[t]
\centering\vspace*{-0mm}
\includegraphics[width=3in]{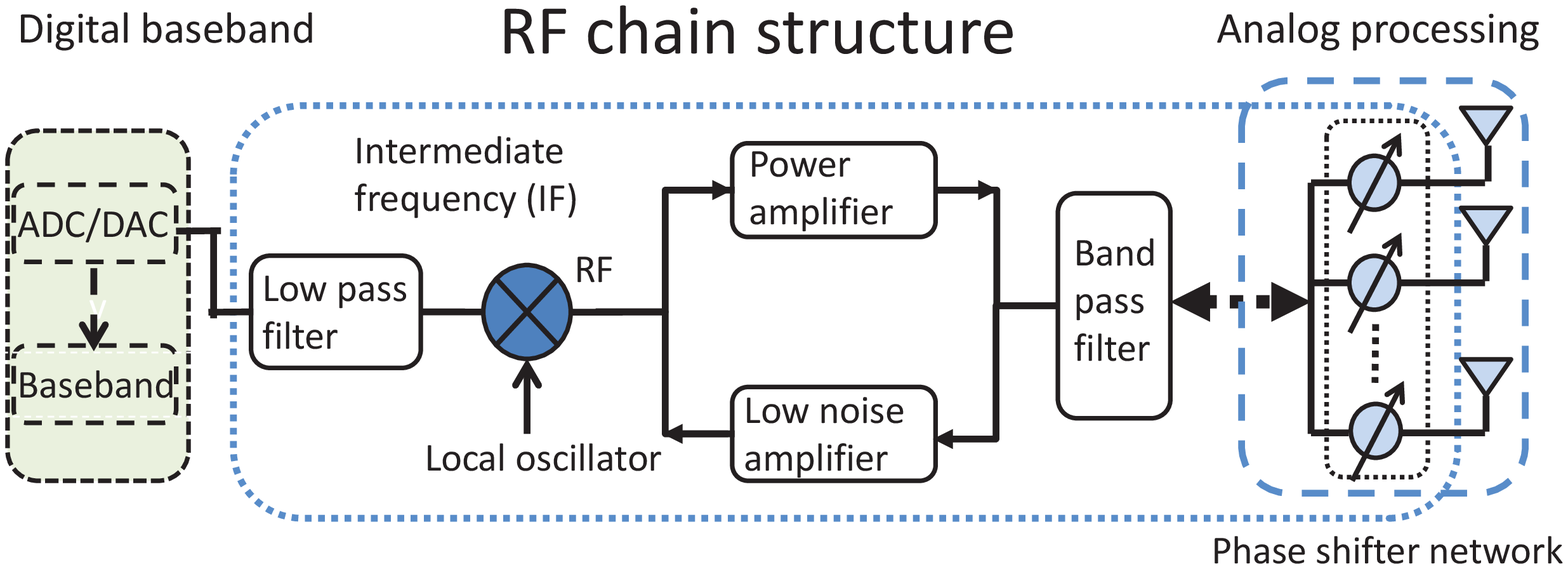}
\vspace{-4mm}
\caption{A block diagram of a RF chain of a subarray antenna structure.}
\label{fig:RF_chain}\vspace{-4mm}
\end{figure}

In this section, we adopt the algorithm proposed in \cite{Zhao2017} for the estimation of an equivalent mmWave channel, which is based on analog beamforming matrices adopted at the desired transceivers and the physical mmWave channels.
The proposed algorithm is suitable for both the conventional fully digital systems and the emerging hybrid systems with fully access and subarray implementation structures.
For the sake of presentation, we provide a brief summary of the algorithm proposed in \cite{Zhao2017} in the following.
For illustration, we adopt the subarray structure as an example, as shown in Fig. \ref{fig:RF_chain}.
Specifically, each RF chain can access to the $M$ antennas via a phase shifter network.

There are three steps in the algorithm proposed in \cite{Zhao2017}.
In first and second step, strongest AoAs between the desired BS and the users are estimated.
Then, relying on the estimated strongest AoAs, the desired BS and the users design their analog beamforming matrices.
In the third step, orthogonal pilot symbols are transmitted from the users to the BS by using the pre-designed analog beamforming matrices.
Subsequently, by exploiting the channel reciprocity, the equivalent downlink channel can be estimated and adopt at the BS as the input for the digital baseband precoder.

Note that, orthogonal pilot symbols are used for the estimation of the equivalent channels in the third step.
For the multi-cell scenario, orthogonal pilot symbols are reused among different cells which results in pilot contamination and cause a severe impact on the equivalent channel estimation performance.
We note that only single-cell scenario was considered in \cite{Zhao2017} and it is unclear if the channel estimation algorithm provides robustness against pilot contamination.
Furthermore, the performance analysis studied in \cite{Zhao2017} does not take into account the impact of potential out-of-cell interference on the performance of channel estimation.
In the following sections, we investigate the impact of pilot contamination on the mmWave channel estimation performance for small-cell scenarios.

\subsection{Channel Estimation and Pilot Contamination Analysis}
\setcounter{equation}{11}
\begin{figure*}[t]
\begin{align}
\hspace{-3mm}\widehat{\mathbf{H}}_{\mathrm{eq}}^{T}=  \mathbf{B}\left(\frac{\mathbf{\Psi }^{H}}{E_\mathrm{P}}\left[\begin{array}{ccc}\widehat{\mathbf{s}}_{1} & \ldots & \widehat{\mathbf{s}}_{N}\end{array}\right]\right)
=\mathbf{B}\underset{\mathbf{H}_{\mathrm{eq}}^{T}}{\underbrace{\left[
\begin{array}{c}
\widehat{\bm{\omega }}_{1}^{H}\mathbf{H}_{1}^{T}\mathbf{F}_{\mathrm{RF}%
} \\
\vdots \\
\widehat{\bm{\omega}}_{N}^{H}\mathbf{H}_{N}^{T}\mathbf{F}_{\mathrm{RF}}\end{array}\right]}}+\underset{\mathrm{Effective}\text{\ }\mathrm{noise}}{\underbrace{\frac{\mathbf{B}}{\sqrt{E_{\mathrm{P}}}}\left[
\begin{array}{c}
\mathbf{\Phi}_{1}^{H}\mathbf{Z}^{T}\mathbf{F}_{\mathrm{RF}} \\
\vdots \\
\mathbf{\Phi} _{N}^{H}\mathbf{Z}^{T}\mathbf{F}_{\mathrm{RF}}
\end{array}\right] }} +
\underset{\mathrm{Pilot}\text{\ }\mathrm{contamination}}{\underbrace{\mathbf{B}\left[
\begin{array}{c}
\overset{L}{\underset{l=1}{\sum }}\left(\widehat{\bm{\omega}}^{H}_{l,1}\Delta \widehat{\mathbf{H}}_{l,1}^{T}\right)\mathbf{F}_{\mathrm{RF}} \\
\vdots \\
\overset{L}{\underset{l=1}{\sum }}\left(\widehat{\bm{\omega}}^{H}_{l,N}\Delta \widehat{\mathbf{H}}_{l,N}^{T}\right)\mathbf{F}_{\mathrm{RF}}
\end{array}\right] }}. \label{Eq_eq}
\end{align}
\hrule\vspace{-5mm}
\end{figure*}

Basically, the first and second step provide the analog beamforming matrices at the desired BS and the desired users to facilitate the estimation of equivalent channel.
In particular, the analog beamforming matrices pair the desired BS and the desired users and align the directions of data stream transmission.
Due to the inter-cell large scale propagation path loss, the impact of pilot contamination on strongest AoAs estimation is usually negligible\footnote{For strongest AoAs estimation at the desired BS in a multi-cell scenario, the received power of the reused pilot symbols transmitted from the users in neighboring cells is smaller than that of the desired pilot symbols transmitted from the users in the desired cell. In addition, the strongest AoAs estimation may not rely on pilot symbols \cite{Zhao2017B}.}.
If readers are interested in the impacts of strongest AoA estimation errors on the equivalent channel estimation performance and the downlink rate performance, please refer to \cite{Zhao2017} for further details.

Therefore, to facilitate the performance analysis of the multi-cell equivalent channel estimation, we assume that strongest AoAs among the users and the BS are perfectly estimated and the desired signals always fall in the mainlobe.
We note that the impact of multi-cell interference on the estimation of strongest AoAs will be captured in the simulation.


Based on the perfectly estimated strongest AoAs at the users and the BS, the analog receive beamforming vector of user $k$ at the desired BS is given by
\setcounter{equation}{7}
\begin{equation}
\widehat{\bm{\gamma}}^{T}_{k}\in\mathbb{C}^{1\times M} =
\frac{1}{\sqrt{M}}\left[
\begin{array}{ccc}
1,  \ldots , e^{j2\pi \left( M-1\right) \tfrac{d}{\lambda }\cos
\left(\theta_{k}\right)}
\end{array}\right]
\end{equation}
and the analog transmit beamformer of user $i$ in the desired cell is given by
\begin{equation}
\widehat{\bm{\omega}}^{\ast}_{i} \in\mathbb{C}^{P\times 1}=
\frac{1}{\sqrt{P}}\left[
\begin{array}{ccc}
1, \ldots , e^{j2\pi \left( P-1\right) \tfrac{d}{\lambda }\cos
\left(\phi_{i}\right) }
\end{array}\right] ^{H}.
\end{equation}
In addition, we denote the designed analog beamforming matrix at the desired BS as
\begin{equation}
\mathbf{F}_{\mathrm{RF}}\in\mathbb{C}^{M\times N}=\left[\begin{array}{ccc}\widehat{\bm{\gamma}}_{1},\ldots, \widehat{\bm{\gamma}}_{N}\end{array}\right].
\end{equation}
Let $\mathbf{\Phi }_{k}\in\mathbb{C}^{N\times 1}$ denote the pilot symbols of user $k$ in the desired cell.

The pilot symbols for all the $N$ users in the desired cell form a matrix, $\mathbf{\Psi \in\mathbb{C}}^{N\times N}\mathbf{,}$ where $\mathbf{\Phi }_{k}$ is a column vector of
matrix $\mathbf{\Psi }$ given by
$\mathbf{\Psi }=\ \sqrt{E_{\mathrm{P}}}\left[
\begin{array}{ccc}
\mathbf{\Phi }_{1},\ldots,\mathbf{\Phi }_{N}%
\end{array}\right]$, $\mathbf{\Phi }_{i}^{H}\mathbf{\Phi }_{j}=0$, $\forall i\neq j$, $i,\text{ }j\in \left\{ 1,\ldots, N\right\}$,
where $E_{\mathrm{P}}$ represents the transmitted pilot symbol energy.
During the channel estimation phase, the reuse of pilot symbols in neighboring cells affects the performance of equivalent channel estimation.
The received signal of the $k$-th RF chain at the desired BS in the uplink is given by\vspace{-2mm}
\begin{align}
\widehat{\mathbf{s}}_{k}^{T}  =&\widehat{\bm{\gamma}}_{k}^{T}\overset{N}{\underset{i=1}{\sum }}\mathbf{H}_{i}\widehat{\bm{\omega }}_{i}^{\ast }\sqrt{E_{%
\mathrm{P}}}\mathbf{\Phi }_{i}^{T}\notag\\
&+\underset{\mathrm{Pilot\text{ \ }contamination}}{\underbrace{\widehat{\bm{\gamma}}_{k}^{T}\overset{L}{\underset{l=1}{\sum }}\overset{N}{\underset{i=1}{\sum }}\left(\Delta \widehat{\mathbf{H}}_{l,i}\widehat{\bm{\omega}}^{\ast}_{l,i}\sqrt{E_{\mathrm{P}}}\mathbf{\Phi }_{i}^{T}\right)}}+\widehat{\bm{\gamma}}_{k}^{T}\mathbf{Z},\vspace{-2mm}
\end{align}
where $\widehat{\bm{\omega}}_{l,i}\in\mathbb{C}^{P\times 1}=
\frac{1}{\sqrt{P}}\left[
\begin{array}{ccc}
1,  \ldots , e^{j2\pi \left( P-1\right) \tfrac{d}{\lambda }\cos
\left(\phi_{l,k}\right) }
\end{array}\right] ^{T}$ is the analog beamforming vector of user $i$ in the $l$-th cell, the entries of noise matrix, $\mathbf{Z}$, are modeled by i.i.d. random variables with distribution $\mathcal{CN}\left( 0,\sigma _{\mathrm{BS}}^{2}\right)$.

To facilitate the investigation of channel estimation and downlink transmission, we assume that long-term power control is performed to compensate different LOS path loss among different desired users in the desired cell.
As a result, it can be considered that the large scale propagation path losses of different users in the desired cell are identical.
Thus, we can express the estimated equivalent downlink channel $\widehat{\mathbf{H}}_{\mathrm{eq}}^{T}\in\mathbb{C}^{N\times N}$ at the desired BS under the impact of pilot contamination in Equation (\ref{Eq_eq}) at the top of this page, where $\Delta \widehat{\mathbf{H}}_{\mathrm{eq}}^{T}$ is the equivalent channel estimation error caused by pilot contamination and noise.
The path loss compensation matrix $\mathbf{B}\in\mathbb{C}^{N\times N}$ is given by \vspace{-1mm}
\setcounter{equation}{12}
\begin{equation}
\mathbf{B}=\left[
\begin{array}{ccc}
\frac{1}{\sqrt{\varpi_{{1}}}} & \cdots &  {0} \\
\vdots & \ddots  & \vdots \\
{0} & \cdots & \frac{1}{\sqrt{\varpi_{{N}}}}%
\end{array}\right].\vspace{-1mm}
\end{equation}
In the following, for notational simplicity, we denote $\widehat{\rho}_{l,k}= \sqrt{\frac{\widehat{\varpi}_{l,{k}}}{\varpi_{{k}}}}$ as the inter-cell propagation path loss coefficients.
Now, to evaluate the impact of pilot contamination, we introduce a theorem which reveals the normalized MSE performance of equivalent channel estimation.

\setcounter{equation}{13}
\begin{theo}
\label{thm:Theo_1}
The normalized MSE of the equivalent channel estimation with respect to the $k$-th RF chain under the impacts of pilot contamination and noise can be approximated as
\begin{align}
\mathrm{NMSE}_{\mathrm{eq,}k} =&\frac{1}{N}
\mathbb{E}_{\Delta\widehat{\mathbf{H}}_{l,i}}\left[ \left( \dfrac{1}{\sqrt{MP}}\Delta\widehat{\mathbf{h}}_{\mathrm{eq},k}^{T}\right)\left(\dfrac{1}{\sqrt{MP}}
\Delta\widehat{\mathbf{h}}_{\mathrm{eq},k}^{\ast}\right) \right]   \notag \\
\approx & \frac{1}{NMP}\left[\overset{L}{\underset{l=1}{\sum }}{\frac{\widehat{\rho} _{l,k}^{2}\upsilon_{l,k}}{\upsilon_{l,k}+1}}N\right]+\frac{1}{MP}\overset{L}{\underset{l=1}{\sum }}{\frac{\widehat{\rho}_{l,k}^{2}}{\upsilon_{l,k}+1}}\notag\\
&+\frac{\sigma _{\mathrm{BS}}^{2}\mathrm{tr}\left[ \mathbf{F}_{\mathrm{RF}%
}^{H}\mathbf{F}_{\mathrm{RF}}\right] }{{\varpi_{{k}}}E_{\mathrm{P}}NMP}\notag\\
= & \underset{\mathrm{Pilot\text{ \ }contamination}}{\underbrace{\frac{1}{MP}\overset{L}{\underset{l=1}{\sum }}\left( \widehat{\rho} _{l,k}^{2} \right)}}+\underset{\mathrm{Noise}}{\underbrace{\frac{\sigma_{\mathrm{BS}}^{2}}{{\varpi_{{k}}}E_{\mathrm{P}}MP}}}, \label{PC}
\end{align}%
In particular, when the number of antennas equipped at the desired BS and the users are sufficiently large, we have
\begin{equation}
\underset{M,P\rightarrow\infty}{\lim}\mathrm{NMSE}_{\mathrm{eq,}k}\approx  0.
\end{equation}
\end{theo}
\emph{\quad Proof: } Please refer to Appendix A.\QEDA

In Equation (\ref{PC}), the impact of the multi-cell pilot contamination term on the normalized MSE performance is inversely proportional to the increasing number of antennas $M$ and $P$.
In addition, the noise term decreases with the increasing transmit pilot symbol energy and the number of antennas, $M$ and $P$.
It is important to note that, although the impact of noise part on channel estimation will vanish in the high SNR regime, e.g. $E_{p}\gg 1 $, it has no influence on the pilot contamination.

It is known that the conventional massive MIMO pilot-aided least-square (LS)  channel estimation performance under the impact of pilot contamination cannot be improved by increasing the number of antennas equipped at the BS \cite{Marzetta2010,Jose2011}.
Interestingly, the result of Theorem $\ref{thm:Theo_1}$ unveils that the impacts of pilot contamination and noise on the equivalent channel estimation will vanish  asymptotically with the increasing number of antennas equipped at each RF chain, $M$ and $P$.
Actually, the numbers of antennas $M$ and $P$ have an identical effect on the normalized MSE performance.
This is because the direction of analog beamforming matrices adopted at the desired BS and the desired users align with the strongest LOS path.
Hence, the analog beamforming matrices adopted at the desired BS and the users form a pair of spatial filters which block the pilot signals from undesired users to the desired BS via non-strongest paths.
In addition, transmitting the pilot signals from the desired users via the analog beamforming matrix can reduce the potential energy leakage to other undesired cells, which further reduces the impact of pilot contamination.
Note that the proposed channel estimation algorithm does not require any information of covariance matrix of the inter-cell channels as required by the MMSE-based precoding algorithm proposed in \cite{Bjornson2017}.
\begin{figure}[t]
\centering\vspace*{-0mm}
{
\includegraphics[width=3.0in,]{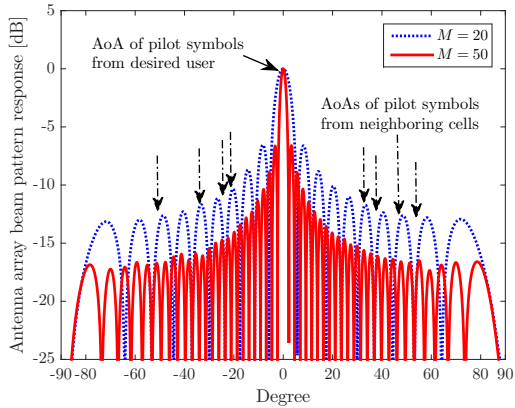}}
\caption{The illustration of sidelobe suppression for different numbers of antennas $M$.}
\label{fig:HBFD}\vspace{-4mm}
\end{figure}

We note that for the conventional pilot-aided channel estimation algorithms, e.g. LS-based algorithms, they estimate the channels from all the directions.
Thus, BSs adopting these algorithms receive reused pilot symbols from the undesired users and cannot be distinguished from the desired pilot symbols.
This is known as pilot contamination.
However, adopting analog beamforming matrices for receiving pilot symbols at the desired BS via strongest AoA directions forms a spatial filter, which blocks the undesired pilot symbols from neighboring cells via different AoA paths.
Furthermore, the ``blocking capability" improves with the increasing number of antennas equipped at the BS.
Specifically, the beamwidth of mainlobe becomes narrower and the magnitude of sidelobes is lower, which is illustrated in Fig. \ref{fig:HBFD}.
In fact, this is an important feature for mitigating the impact of pilot contamination.
Therefore, with the equivalent channel estimation proposed in \cite{Zhao2017}, the impact of pilot contamination vanishes asymptotically with an increasing number of antennas $M$ equipped at the desired BS.
\begin{rema}
Since hybrid mmWave systems are the generalization of fully digital systems, the algorithm proposed in \cite{Zhao2017} can be extended to the case of fully digital systems. In particular, the derived analysis in the paper can be directly applied to the latter systems.
\end{rema}
\vspace{-3mm}
\section{Simulation}

\begin{figure}[t]
\centering{\vspace{-0mm}}
\includegraphics[width=3.0in]{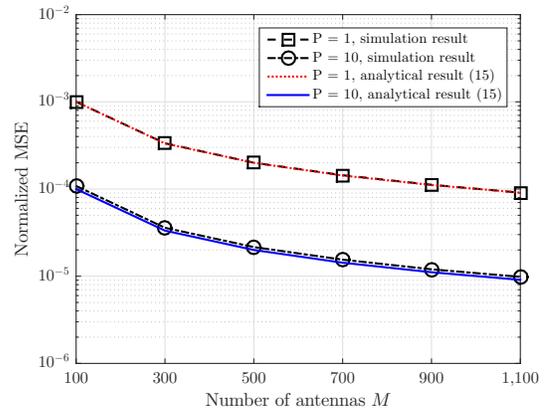}
\caption{The illustration of multi-cell normalized MSE performance under the impact of pilot contamination versus the number of antennas $M$ in the high SNR regime, i.e., maximum transmit power $46$ dB, for $N_{\mathrm{RF}}=N=10$, Rician K-factor $=5$, and channel estimation error $\xi^{2} = 0.01$.}{\vspace{-3mm}}
\label{fig:CE_PC}
\end{figure}

To verify the correctness of analytical results derived in Equation (\ref{PC}), here, we provide some simulation results.
We assume that the antenna gain in zenith is $14$ dBi.
In particular, we focus on the impact of pilot contamination in the high SNR regime and take into account the inter-cell propagation path loss.
The maximum BS transmit power is set as $46$ dBm.
In the simulation, we take into account any possible estimation errors in estimating strongest AoA paths, which verifies the assumption of perfect strongest AoAs estimation adopted for the design of $\mathbf{F}_{\mathrm{RF}}$.
We set the carrier frequency as $30$ GHz, the number of neighboring cells as $L=6$ and the number of users per cell $N=10$.
Fig. \ref{fig:CE_PC} illustrates the normalized MSE of the equivalent channel estimation versus the number of antennas equipped at the desired BS under the impact of pilot contamination.

\begin{figure}[t]
\centering{\vspace{-0mm}}
\includegraphics[width=3.0in]{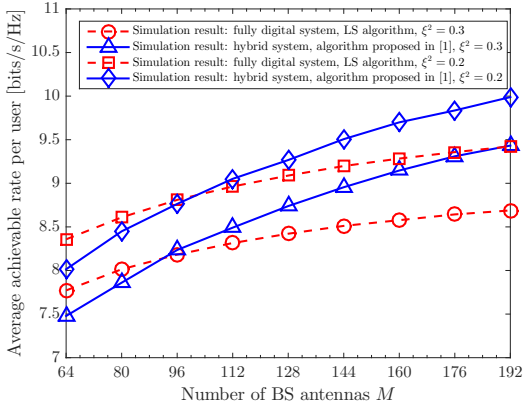}
\caption{The average achievable rate per user (bits/s/Hz) under the impact of pilot contamination versus the number of antennas $M$ in the high SNR regime.}{\vspace{-4mm}}
\label{fig:CE_PC1}
\end{figure}
We note here, while the number of antennas increases, the number of RF chains equipped at the desired BS remains the same.
The channel estimation errors of a user caused by pilot contamination, $\xi^{2}=\overset{L}{\underset{l=1}{\sum }}\left( \widehat{\rho} _{l,k}^{2} \right)$, $\forall k\left\{ 1,\ldots, N\right\}$, is set as $\xi^{2}=0.01$.
In Fig. \ref{fig:CE_PC}, we can observe that with an increasing number of antennas equipped at the desired BS $M$, the normalized MSE decreases monotonically.
In addition, an increasing number of antennas equipped at the users $P$ can also improve the normalized MSE.
Besides, the simulation results match with the analytical results derived in Equation (\ref{PC}).
Thus, the impact of pilot contamination on the channel estimation of multi-cell hybrid mmWave systems vanishes asymptotically, for sufficiently large number of antennas equipped at the desired BS $M$.
In practice, to meet a certain required normalized MSE of channel estimation, we can either increase the number of antennas equipped at each RF chain of the BS or the number of antennas at each user for a fixed pilot symbol energy $E_{p}$.

In Fig. \ref{fig:CE_PC1}, we study the average achievable rate per user under the impact of pilot contamination.
In particular, we adopt ZF precoding strategy for the downlink transmission based on the estimated CSI.
We assume that the channel estimation errors of a user, $\xi^{2}=\overset{L}{\underset{l=1}{\sum }}\left( \widehat{\rho} _{l,k}^{2} \right)$, $\forall k\left\{ 1,\ldots, N\right\}$, at the desired BS due to the impact of pilot contamination effects for both the fully digital system and the hybrid system are set as $\xi^{2} = 0.2$ and $\xi^{2}=0.3$, respectively.
For comparison, we also simulate the performance of a fully digital system adopting conventional LS-based CSI estimation algorithm and ZF for downlink transmission. It can be observed that the average achievable rate of the fully digital system adopting the conventional LS-based CSI estimation algorithm under the impact of pilot contamination is saturated with the increasing number of antennas equipped at the desired BS. In contrast,  the average achievable rate of the hybrid system can scale with the number of BS antennas $M$ as the impact of pilot contamination is mitigated when the numbers of antennas are sufficiently large.

\section{Conclusions}
In this paper, we investigated the normalized MSE performance of the channel estimation proposed in \cite{Zhao2017} for multi-cell hybrid mmWave systems.
The derived closed-form approximation of the normalized channel estimation MSE performance revealed that the channel estimation error caused by the impact of pilot contamination and noise would vanish asymptotically with the increasing number of antennas.
Furthermore, based on the estimated CSI, we adopted ZF precoding for the downlink transmission.
Our simulation results unveiled that the average achievable rate per user  increases with the increasing number of antennas equipped at the BS, despite the impact of pilot contamination. It is an excellent feature for multi-cell hybrid mmWave systems with small-cell radius for improving the spectral efficiency.
\subsection{Proof of Theorem 1}
The MSE performance of equivalent channel estimation under the impact of pilot contamination is given by
\begin{align}
\mathrm{NMSE}_{\mathrm{eq,}k} =&\frac{1}{NMP}
\mathbb{E}_{\Delta\widehat{\mathbf{H}}_{l,i}}\left[ \left( \Delta\widehat{\mathbf{h}}_{\mathrm{eq},k}^{T}\right)\left(
\Delta\widehat{\mathbf{h}}_{\mathrm{eq},k}^{\ast}\right) \right]. \label{Eq_AP1}
\end{align}%
Due to the small radius of cell, the inter-cell uplink propagation channels may contain LOS components from the users in the neighboring cells to the desired BS.
Thus, the part associated with pilot contamination can be expressed as\vspace{-1mm}
\begin{align}
&\hspace{-0mm}\frac{1}{{\varpi_{ {k}}}}\mathbb{E}_{\Delta\widehat{\mathbf{H}}_{l,k}}\left[\overset{L}{\underset{l=1}{\sum }}\left(\widehat{\bm{\omega}}^{H}_{l,k}\Delta \widehat{\mathbf{H}}_{l,k}^{T}\mathbf{F}_{\mathrm{RF}}\mathbf{F}_{\mathrm{RF}}^{H}\Delta \widehat{\mathbf{H}}_{l,k}^{\ast}\widehat{\bm{\omega}}_{l,k}\right)\right] \notag\\
\hspace{-2mm}\approx & {{\mathbb{E}_{\Delta\widehat{\mathbf{H}}_{\mathrm{L},l,k}}\underset{\mathrm{Inter-cell}\text{ }\mathrm{interference}\text{ }\mathrm{caused}\text{ }\mathrm{by}\text{ }\mathrm{LOS}\text{ }\mathrm{component}}{\underbrace{\left[\overset{L}{\underset{l=1}{\sum }}\left({\frac{\widehat{\rho} _{l,k}^{2}\upsilon_{l,k}}{\upsilon_{l,k}+1}}\widehat{\bm{\omega}}^{H}_{l,k}\Delta \widehat{\mathbf{H}}_{\mathrm{L},l,k}^{T}\mathbf{F}_{\mathrm{RF}}\mathbf{F}_{\mathrm{RF}}^{H}\Delta \widehat{\mathbf{H}}_{\mathrm{L},l,k}^{\ast}\widehat{\bm{\omega}}_{l,k}\right)\right]}}}}\notag\\
\hspace{-2mm}+&
{\mathbb{E}_{\Delta\widehat{\mathbf{H}}_{\mathrm{S},l,k}}\underset{\mathrm{Inter-cell}\text{ }\mathrm{interference}\text{ }\mathrm{caused}\text{ }\mathrm{by}\text{ }\mathrm{scattering}\text{ }\mathrm{component}}{\underbrace{\left[\overset{L}{\underset{l=1}{\sum }}\left({\frac{\widehat{\rho} _{l,k}^{2}}{\upsilon_{l,k}+1}}\widehat{\bm{\omega}}^{H}_{l,k}\Delta \widehat{\mathbf{H}}_{\mathrm{S},l,k}^{T}\mathbf{F}_{\mathrm{RF}}\mathbf{F}_{\mathrm{RF}}^{H}\Delta \widehat{\mathbf{H}}_{\mathrm{S},l,k}^{\ast}\widehat{\bm{\omega}}_{l,k}\right)\right]}}},\vspace{-4mm}
\end{align}%
where $\widehat{\rho}_{l,k}= \sqrt{\dfrac{\widehat{\varpi}_{l,{k}}}{\varpi_{{k}}}}$.
Then, the inter-cell interference caused by scattering component can be further approximated as
\begin{align}
&\hspace{-2mm}{{\left[\overset{L}{\underset{l=1}{\sum }}{\frac{\widehat{\rho} _{l,k}^{2}}{\upsilon_{l,k}+1}}\widehat{\bm{\omega}}^{H}_{l,k}\mathbb{E}_{\Delta\widehat{\mathbf{H}}_{\mathrm{S},l,k}}\left(\Delta \widehat{\mathbf{H}}_{\mathrm{S},l,k}^{T}\mathbf{F}_{\mathrm{RF}}\mathbf{F}_{\mathrm{RF}}^{H}\Delta \widehat{\mathbf{H}}_{\mathrm{S},l,k}^{\ast}\right)\widehat{\bm{\omega}}_{l,k}\right]}}\notag\\
\hspace{-2mm}\approx &\hspace{-0mm}\overset{L}{\underset{l=1}{\sum }}{\frac{\widehat{\rho} _{l,k}^{2}}{\upsilon_{l,k}+1}}\widehat{\bm{\omega}}^{H}_{l,k}{\mathrm{tr}}\left(\mathbf{F}_{\mathrm{RF}}\mathbf{F}_{\mathrm{RF}}^{H}\right)\mathbf{I}_{\mathrm{P}}\widehat{\bm{\omega}}_{l,k}
\hspace{-1mm}=\hspace{-1mm}\overset{L}{\underset{l=1}{\sum }}{\frac{\widehat{\rho} _{l,k}^{2}}{\upsilon_{l,k}+1}}N. \label{Eq_AP2}
\end{align}
Now, we would like to approximate the inter-cell interference caused by multi-cell LOS components
\begin{align}
&\hspace{-1mm}\mathbb{E}_{\Delta\widehat{\mathbf{H}}_{\mathrm{L},l,k}}\left[\overset{L}{\underset{l=1}{\sum }}\left({\frac{\widehat{\rho} _{l,k}^{2}\upsilon_{l,k}}{\upsilon_{l,k}+1}}\widehat{\bm{\omega}}^{H}_{l,k}\Delta \widehat{\mathbf{H}}_{\mathrm{L},l,k}^{T}\mathbf{F}_{\mathrm{RF}}\mathbf{F}_{\mathrm{RF}}^{H}\Delta \widehat{\mathbf{H}}_{\mathrm{L},l,k}^{\ast}\widehat{\bm{\omega}}_{l,k}\right)\right] \notag \\
&\hspace{-2mm}=\mathbb{E}_{\Delta\widehat{\mathbf{H}}_{\mathrm{L},l,k}}\left[\overset{L}{\underset{l=1}{\sum }}\mathrm{tr}\left[{\frac{\widehat{\rho} _{l,k}^{2}\upsilon_{l,k}}{\upsilon_{l,k}+1}}\left(\Delta\widehat{\mathbf{h}}_{\mathrm{L,}l,k}^{\mathrm{BS}}\right)^{T}\mathbf{F}_{\mathrm{RF}}
\mathbf{F}_{\mathrm{RF}}^{H}\left(\Delta\widehat{\mathbf{h}}_{\mathrm{L,}l,k}^{\mathrm{BS}}\right)^{\ast}\right.\right.\notag \\
&\text{\ \ \ \ \ \ \ \ \ \ \ \ \ \ \ \ \ \ \ \ \ \ \ }\left.\left.\Delta\widehat{\mathbf{h}}_{%
\mathrm{L,}l,k}^{T}\widehat{\bm{\omega}}_{l,k}\widehat{\bm{\omega}}^{H}_{l,k}\Delta\widehat{\mathbf{h}}_{%
\mathrm{L,}l,k}^{\ast}\right]\right]. \label{Eq_apdix1}
\end{align}%
In Equation (\ref{Eq_apdix1}), we have
\begin{align}
\hspace{-3mm}\widehat{\bm{\omega }}_{l,k}^{H}&=\frac{1}{\sqrt{P}}\left[
\begin{array}{ccc}
1, & \ldots , & e^{j2\pi \left( P-1\right) \tfrac{d}{\lambda }\cos \left({\phi}_{l,k}\right) }%
\end{array}%
\right] ^{\ast} \text{and} \\
\hspace{-3mm}\Delta \widehat{\mathbf{h}}_{\mathrm{L},l,k}^{\ast} &=\left[\begin{array}{ccc} 1, & \ldots , & \text{ }e^{-j2\pi \left( P-1\right) \tfrac{d}{\lambda }\cos \left( \Delta\phi_{l,k}\right) }\end{array}\right] ^{H},
\end{align}%
where variables $\phi _{l,k}\in \left[0, \text{\ }\pi \right]$ is the angle of incidence of the LOS path at antenna arrays of user $k$ in cell $l$, and $\Delta\phi _{l,k}\in \left[0, \text{\ }\pi \right] $ is the angle of incidence of the inter-cell LOS path at antenna arrays from user $k$ of cell $l$ to the desired BS.
By defining the array gain function $G_{\mathrm{act},P}{\left[x\right]}$, cf. \cite{book:wireless_comm}, as\vspace*{-1mm}
\begin{equation}
G_{\mathrm{act},P}{\left[x \right]} = \frac{\left\{\sin\left[ P \pi\frac{d}{\lambda}\left( x \right)\right]\right\}^{2}}{{P}\left\{\sin\left[\pi\frac{d}{\lambda}\left( x \right)\right]\right\}^{2}},\vspace*{-2mm}
\end{equation}where $\dfrac{d}{\lambda}=\dfrac{1}{2}$.
Then, we have:
\begin{align}
&\hspace*{-1mm}G_{\mathrm{act},P}{\left[ \cos\left(\phi_{l,k}\right) - \cos\left(\Delta\phi_{l,k}\right) \right]}=\Delta\widehat{\mathbf{h}}_{%
\mathrm{L,}l,k}^{T}\widehat{\bm{\omega}}_{l,k}\widehat{\bm{\omega }}_{l,k}^{H}\Delta\widehat{\mathbf{h}}_{\mathrm{L},l,k}^{\ast}\notag\\
=&\frac{\left\{\sin\left[ P \pi\frac{1}{2}\left( \cos \left(\phi_{l,k}\right)-\cos\left(\Delta\phi_{l,k}\right)   \right)\right]\right\}^{2}}{{P}\left\{\sin\left[\pi\frac{1}{2}\left( \cos \left(\phi_{l,k}\right)-\cos\left(\Delta\phi_{l,k}\right)   \right)\right]\right\}^{2}} . \label{Eq_AP3}
\end{align}
It is also known that $\cos\left(\phi_{l,k}\right)$ and $\cos\left(\Delta\phi_{l,k}\right)$ are independent uniformly distributed over $[-1, \text{\ } 1]$.
Due to the periodic property of function $e^{j2\pi x}$, the linear antenna array gain $G_{\mathrm{act},P}{\left[ \cos\left(\phi_{l,k}\right) - \cos\left(\Delta\phi_{l,k}\right) \right]}$ is equal in distribution to $G_{\mathrm{act},P}{\left[\mu_{l,k}\right]}$, where $\mu_{l,k}$, $\forall k \in \left\{ 1,\ldots,N \right\}$, is uniformly distributed over $[-1, \text{\ } 1]$ (Lemma $1$ of \cite{Yu2017}).
Similarly, we can have following preliminaries, i.e.,
\begin{align}
\hspace{-3mm}\left(\Delta\widehat{\mathbf{h}}_{\mathrm{L},l,k}^{\mathrm{BS}}\right)^{T}&=\left[\begin{array}{ccc}1,\ldots,\text{ }e^{-j2\pi \left( M-1\right) \tfrac{d}{\lambda }\cos \left(\theta _{l,k}\right) }\end{array}\right] \text{\ and}\notag\\
\hspace{-3mm}\widehat{\bm{\gamma}}_{i}&=
\frac{1}{\sqrt{M}}\left[
\begin{array}{ccc}
1,\ldots ,\text{ }e^{j2\pi \left( M-1\right) \tfrac{d}{\lambda }\cos
\left( {\theta }_{i}\right) }
\end{array}%
\right]^{T}.
\end{align}
Based on these aforementioned expressions, we rewrite $\left(\Delta\widehat{\mathbf{h}}_{\mathrm{L,}l,k}^{\mathrm{BS}}\right)^{T}\mathbf{F}_{\mathrm{RF}}
\mathbf{F}_{\mathrm{RF}}^{H}\left(\Delta\widehat{\mathbf{h}}_{\mathrm{L,}l,k}^{\mathrm{BS}}\right)^{\ast}$
as
\begin{align}
&\left[\begin{array}{ccc}\left(\Delta\widehat{\mathbf{h}}_{\mathrm{L,}l,k}^{\mathrm{BS}}\right)^{T}\widehat{\bm{\gamma}}_{1},\ldots, \left(\Delta\widehat{\mathbf{h}}_{\mathrm{L,}l,k}^{\mathrm{BS}}\right)^{T}\widehat{\bm{\gamma}}_{N}\end{array}\right]\notag\\
&\times\left[\begin{array}{ccc}\widehat{\bm{\gamma}}^{H}_{1}\left(\Delta\widehat{\mathbf{h}}_{\mathrm{L,}l,k}^{\mathrm{BS}}\right)^{\ast},\ldots, \widehat{\bm{\gamma}}^{H}_{N}\left(\Delta\widehat{\mathbf{h}}_{\mathrm{L,}l,k}^{\mathrm{BS}}\right)^{\ast}\end{array}\right]^{T}\notag\\
&\hspace{-5mm}=G_{\mathrm{act},M}{\left[ \cos\left(\phi_{l,k}\right) - \cos\left(\Delta\phi_{l,k}\right) \right]}, \label{Eq_AP4}
\end{align}%
$G_{\mathrm{act},M}{\left[ \cos \left( \theta _{l,k}\right)  - \cos \left( {\theta }_{i}\right) \right]}$ is equal in distribution to $G_{\mathrm{act},M}{\left[  \epsilon_{l,k,i} \right]}$, where $\epsilon_{l,k,i}$, $\forall i \in \left\{ 1,\ldots,N \right\}$, is also independent uniformly distributed over $[-1, \text{ \ } 1]$.
Substituting Equations (\ref{Eq_AP3}) and (\ref{Eq_AP4}) into Equation (\ref{Eq_apdix1}), we have:
\begin{align}
&\overset{L}{\underset{l=1}{\sum }}\left\{{\frac{\widehat{\rho} _{l,k}^{2}\upsilon_{l,k}}{\upsilon_{l,k}+1}}\mathbb{E}_{\phi_{l,k}, \phi_{l,k} }\left[G_{\mathrm{act},P}{\left[ \cos\left(\phi_{l,k}\right) - \cos\left(\Delta\phi_{l,k}\right) \right]}\right]\right.\notag\\
&\left.\times\left[\overset{N}{\underset{i=1}{\sum }}\mathbb{E}_{\theta _{l,k},\theta_{i} }\left[G_{\mathrm{act},M}{\left[ \cos \left( \theta _{l,k}\right)  - \cos \left( {\theta }_{i}\right) \right]} \right]\right]\right\}\notag\\
=&\mathbb{E}_{\mu_{l,k},{\epsilon}_{l,k,i}}\left[\overset{L}{\underset{l=1}{\sum }}{\frac{\widehat{\rho} _{l,k}^{2}\upsilon_{l,k}}{\upsilon_{l,k}+1}}\left(\frac{\left[\sinc{\left( \frac{\pi}{2}P\mu_{l,k}\right)}\right]^{2}P}{\left[\sinc{\left( \frac{\pi }{2}\mu_{l,k}\right)}\right]^{2}}\right)\right.\notag\\
&\left.\times \left(\overset{N}{\underset{i=1}{\sum }}\frac{\left[\sinc{\left( \frac{\pi }{2}M\epsilon_{l,k,i}\right)}\right]^{2}M}{\left[\sinc{\left( \frac{\pi }{2}\epsilon_{l,k,i}\right)}\right]^{2}}\right)\right]\notag \\
\hspace*{-4mm}\overset{(b)}{\geqslant} & \overset{L}{\underset{l=1}{\sum }}{\frac{\widehat{\rho} _{l,k}^{2}\upsilon_{l,k}}{\upsilon_{l,k}+1}}\mathbb{E}_{\mu_{l,k}}\left[{\left[\sinc{\left( \frac{\pi }{2}P\mu_{l,k}\right)}\right]^{2}P}\right]\notag\\
&\hspace*{-4mm}\times \mathbb{E}_{{\epsilon}_{l,k,i}}\hspace*{-1mm}\left[\overset{N}{\underset{i=1}{\sum }}{\left[\sinc{\left( \frac{\pi }{2}M\epsilon_{l,k,i}\right)}\right]^{2}\hspace*{-1mm}M}\right]
\hspace*{-1mm}\overset{(c)}{\approx}\hspace*{-1mm}  \overset{L}{\underset{l=1}{\sum }}{\frac{\widehat{\rho} _{l,k}^{2}\upsilon_{l,k}}{\upsilon_{l,k}+1}}N. \label{Eq_AP5}
\end{align}
In (b), we exploit the fact that\vspace*{-2mm}
\begin{equation}
[\sinc(x)]^{2} = \left(\frac{\sin{x}}{x}\right)^2\leqslant 1.\vspace*{-1mm}
\end{equation}%
In (c), we explore the law of integration of sinc function for the number of antennas $M$ is sufficiently large, i.e.,\vspace*{-1mm}
\begin{align}
\hspace*{-1mm}\mathbb{E}_{{\epsilon}_{l,k,i}}\left[{\left[\sinc{\left( \dfrac{\pi }{2}M\epsilon_{l,k,i}\right)}\right]^{2}}M\right] \hspace*{-1mm}
\overset{M\rightarrow\infty}\approx \hspace*{-1mm}&\frac{1}{\pi}\int^{\infty}_{-\infty}{\left[\sinc{\left(\chi\right)}\right]^{2}}d\chi =\hspace*{-1mm} 1,
\end{align}
where $\chi=\dfrac{\pi}{2}M\epsilon_{l,k,i}$.
We substitute Equation (\ref{Eq_AP2}) and (\ref{Eq_AP5}) into (\ref{Eq_AP1}), the expression in (\ref{PC}) comes immediately after some straightforward mathematical manipulation.

\vspace*{-0mm}

\bibliographystyle{IEEEtran}
\bibliography{L_Z}

\end{document}